\documentclass[a4paper,11pt]{article}
\pdfoutput=1
cm \voffset = -2truecm

\usepackage{graphicx}
\usepackage{amsmath}
\usepackage{amssymb}
\usepackage{latexsym}
\usepackage[colorlinks]{hyperref}
\usepackage{color}
\usepackage{cite}
\usepackage{makeidx}
\usepackage{float}
\usepackage{multirow}

\restylefloat{figure}

\newcommand{\bra}{\begin{array}}
\newcommand{\era}{\end{array}}
\newcommand{\beq}{\begin{equation}}
\newcommand{\eeq}{\end{equation}}
\newcommand{\bqr}{\begin{eqnarray}}
\newcommand{\eqr}{\end{eqnarray}}

\def\BC{\bb C}
\def\_\BC{\bbi C}

\def\no2 {{\textstyle{n\over 2}}}

\newcommand{\da}{\dagger}

\newcommand{\lb}{\label}

\begin{document}
\begin{titlepage}
\setcounter{page}{1}
\renewcommand{\thefootnote}{\fnsymbol{footnote}}

\begin{flushright}
\end{flushright}

\vspace{5mm}
\begin{center}

{\Large \bf {Double Barriers and Magnetic Field in Bilayer Graphene}}

\vspace{5mm}
{\bf Ilham Redouani}$^{a}$,
{\bf Ahmed Jellal\footnote{\sf ajellal@ictp.it --
a.jellal@ucd.ac.ma}}$^{a,b}$ and {\bf Hocine Bahlouli}$^{c}$

\vspace{5mm}

{$^{a}$\em Theoretical Physics Group,  
Faculty of Sciences, Choua\"ib Doukkali University},\\
{\em PO Box 20, 24000 El Jadida, Morocco}

{$^b$\em Saudi Center for Theoretical Physics, Dhahran, Saudi
Arabia}


{$^c$\em Physics Department, King Fahd University of Petroleum and Minerals,\\
Dhahran 31261, Saudi Arabia}


\vspace{3cm}

\begin{abstract}

We study the transmission probability in an AB-stacked bilayer
graphene of Dirac fermions scattered by
a double barrier structure in the presence of a magnetic field. We
take into account the full four bands of the energy spectrum and use the boundary
conditions to determine the transmission probability. Our numerical results
show that for energies higher than the interlayer coupling,
four 
ways for transmission probabilities are possible
while for energies less than the height of the barrier,
Dirac fermions exhibits transmission resonances and only one
transmission channel is available.
We show that, for AB-stacked bilayer graphene, there is no Klein
tunneling at normal incident.
We find that the transmission displays sharp peaks inside the transmission gap around the Dirac
point within the barrier regions while they are absent around the Dirac point in the well region.
The effect of the magnetic field, interlayer electrostatic
potential and  various barrier geometry parameters on the
transmission probabilities are also discussed.

\end{abstract}
\end{center}

\vspace{3cm}

\noindent PACS numbers: 73.22.Pr, 72.80.Vp, 73.63.-b

\noindent Keywords: Bilayer graphene, double barriers, magnetic
field, transmission.
\end{titlepage}


\section{Introduction}

Graphene \cite{Geim07} is a single layer of carbon atoms packed in a
hexagonal Bravais lattice. This special atomic arrangement gives graphene
truly unique and remarkable physical properties. Its thermal conductivity
is 15 times larger than that of copper and its electron mobility is 20
times larger than that of GaAs. In addition, graphene is a transparent conductor
and has peculiar electronic properties, such as an unusual quantum Hall effect
\cite{Novo05,Zhang05} and its conductivity can be modified over a wide range of values
either by chemical doping or by applying an electric field. In fact it is the
very high mobility of graphene \cite{Moro08} which makes the
material very interesting for electronic high speed applications \cite{Lin10}.
From the theoretical point of view most of these unusual electronic properties have been
associated with the fact that current carrier in graphene are described in terms
of massless two-dimensional Dirac particles \cite{Castro09}.

For practical applications it has been realized that layered graphene system can play
an important role. Hence a few-layer graphene system can be constructed by stacking graphene sheets
on top of each other to form a layered graphene. Bilayer graphene, which is of interest to us in this work,
is a system consisting of two coupled monolayers of carbon atoms,
each with a honeycomb crystal structure \cite{McCa06,novo06}.
Bilayer graphene has many of the properties that are similar to
those of monolayer \cite{Castro09,novo12}.
For monolayer graphene, one can create an energy gap in the
spectrum in many different ways, such as by coupling to substrate
or doping with impurities \cite{Zhou07,Costa07}, while in bilayer
graphene by applying an external electric field
\cite{Zhang09,McCanB09}. In addition, monolayer graphene has a
linear electronic spectrum in the vicinity of the Dirac points.
 However bilayer graphene has four bands where the lowest
conduction band and highest valence band have quadratic spectra,
each pair is separated by an interlayer coupling energy of order
$\gamma_1=0.4\ eV$ \cite{Guinea06,Latil06,Parto06}. It is
well-known that, in the case of monolayer graphene, the
electrostatic potential barriers are fully transparent for low
energy Dirac fermions at normal incidence, which is referred to as
Klein tunneling \cite{Katsn06} but for bilayer graphene, no Klein
tunneling is expected.

Based on previous investigations of Dirac fermions in bilayer
graphene and in particular the work \cite{Duppen} 
in our recent
work \cite{Hassan,Jellal} 
we developed a theoretical
framework to deal with bilayer graphene in the presence of a
perpendicular electric and magnetic fields for single barrier. Our
theoretical model is based on the well established tight binding
Hamiltonian of graphite \cite{Wallace} and adopted the
Slonczewski-Weiss-McClure parametrization of the relevant
intralayer and interlayer couplings \cite{McClure} to model the
bilayer graphene system. In the present work, we study the
transmission probabilities in AB-stacked bilayer graphene by
considering Dirac fermions scattered by double barrier structure
in the presence of a magnetic field. By requiring the continuity
of the wave functions at interfaces, we find the transmission
probabilities. Systematic study revealed that interlayer
interaction is essential, in particular the direct interlayer
coupling parameter $\gamma_1$, for the study of transmission
properties. For energies higher than the interlayer coupling, $E
> \gamma_1$, two propagation modes are available for transport, four possible ways
for transmission probabilities are available. While, when the
energy is less then the height of the barrier, $E < \gamma_1$, the
Dirac fermions exhibits transmission resonances and only one mode
of propagation is available. This work allowed us to compare our
numerical results with existing literature on the subject.

The present paper is organized as follows. In section 2, we
formulate our model by setting the Hamiltonian system and
determining the associated energy eigenvalues in each potential
region. Then, we consider the five potential regions one at a
time, we obtain the spinor solution corresponding to each region
in terms of barrier parameters and applied fields. In section 3,  we use the
boundary conditions to calculate the transmission probabilities in terms of different physical parameters.
In section 4, we present our numerical results for the transmission
probabilities for two cases: when the incident electron energy is
either smaller or greater than the interlayer coupling parameter.
Finally 
 we conclude our work and discuss its potential importance.

\section{Energy spectrum}

\begin{figure}[H]
 \centering
\includegraphics[width=9.5cm, height=4cm]{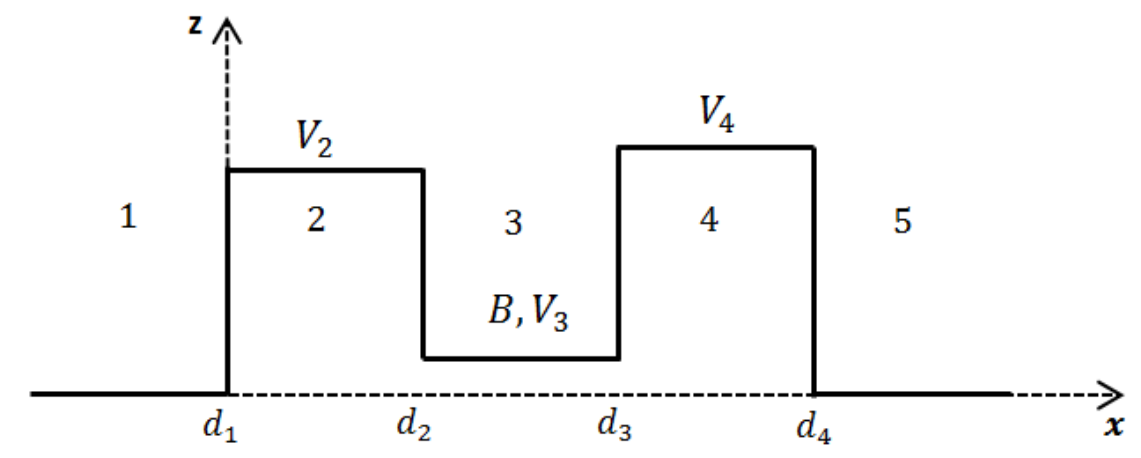}
  \caption{\sf {Schematic diagram for the double barrier, on the first and second layer, in the presence of a magnetic field.}}\label{structuredeV.}
\end{figure}

The considered bilayer graphene system in the presence a perpendicular
electric and magnetic fields is shown in Figure
\ref{structuredeV.}. The Dirac fermions are scattered by a double
barrier potential along the \textit{x}-direction which results in
five different regions. Therefore, the charge carriers in bilayer
AB-stacked are described, in each region denoted by $j$ ($j=1, 2, 3, 4,
5$), by the following four-band Hamiltonian \cite{Castro09,Snyman} and
the associated eigenspinor $\psi(x,y)$ 
\begin{equation}\label{effective hamiltonien}
H_j=\left(%
\begin{array}{cccc}
  V^{+}_{j} & v_F \pi_{j}^{+} & 0 & 0 \\
  v_F \pi_{j} &  V^{+}_{j}  & \gamma_1 & 0 \\
  0 & \gamma_1 &  V^{-}_{j}  & v_F \pi_{j}^{+} \\
   0 & 0 & v_F \pi_{j} &  V^{-}_{j}  \\
\end{array}%
\right), \qquad \psi(x,y)=\left(%
\begin{array}{cccc}
  \psi_{A_1}(x,y) \\
   \psi_{B_1}(x,y) \\
   \psi_{A_2}(x,y)  \\
  \psi_{B_2}(x,y) \\
\end{array}%
\right)
\end{equation}
where $\pi_{j}=p_{x}+ip_{y}$, $p_{x,y}=-i\hbar \nabla+e A_j(x,y)$
is the in-plane momentum relative to the Dirac point, $v_F=
10^{6} m/s$ is the Fermi velocity.
 $V^{+}_{j}$ and $V^{-}_{j}$ are the potentials on the first and
second layer as defined below
\begin{equation} \label{eq 2}
V^{\tau}_{j}=\left\{\begin{array}{ll}
{V_{j}+\tau\delta_{j},}  & {j=2,3,4} \\
{0,}  & {j=1,5}
\end{array}\right.
\end{equation}
in the $j$-th region as shown in Figure
\ref{structuredeV.},
 with $\tau= +1$ ($-1$) on the first layer (second layer), $V_j$
is the barrier strength and $\delta_j$ is the interlayer electrostatic potential
difference.

\noindent The system is considered to be infinite along the \textit{y}-direction.
In region $j=3$, the magnetic field is chosen to be
perpendicular to the graphene sheet along the
\textit{z}-direction and defined as
\beq
B(x,y)=B\Theta
\left[(d_2-x)(d_3-x)\right]
\eeq
 where $B$ is constant and $\Theta$
is the Heaviside step function. In the Landau gauge, the
corresponding vector potential $A(x,y)=(0,A_{y}(x))$
takes the form
\begin{equation}\label{vector potential}
A_{y}(x)=\frac{\hbar}{e l_{B}^2}\left\{\begin{array}{lllll}
{d_{2}}, & \mbox{if} & {x < d_{2}}\\
{x}, & \mbox{if} & {d_{2} < x < d_{3}}\\
{d_{3}}, & \mbox{if} & {x > d_{4}}\\
\end{array}\right.
\end{equation}
 $l_B =\sqrt{\hbar/e B}$ is the magnetic length and $e$
the electronic charge.

In order to solve the eigenvalue problem we
can separate variables and write the eigenspinors as plane waves
in the \textit{y}-direction. This is due to the fact that $[H,p_y]=0$
requires the conservation of momentum along the
\textit{y}-direction, then we can write
$\phi(x,y)=e^{ik_yy}\psi(x,k_y)$.
Using the eigenvalue equation $H\psi(x,k_y)=E\psi(x,k_y)$, we
obtain the following second-order differential equation for
$\psi_{B_1}(x,k_y)$
\begin{equation}\label{eqgnrl}
\left[2\vartheta_{0}^2aa^{\da}-\left(E-V_{j}+\delta_{j}\right)^{2}\right]\left[2\vartheta_{0}^2
a^{\da}a-\left(E-V_{j}-\delta_{j}\right)^{2}\right]\psi_{B_1}
=\gamma_{1}^{2}\left[\left(E-V_{j}\right)^2-\delta_{j}^2\right]\psi_{B_1}
\end{equation}
where $\vartheta_0=\frac{\hbar v_F}{l_B}$ is the energy scale. We have introduced the annihilation and creation operators
\beq
a=\frac{l_B}{\sqrt{2}}\left(\partial_x+k_y+\frac{e}{\hbar}A_y(x)\right), \qquad
a^{\da}=\frac{l_B}{\sqrt{2}}\left(-\partial_x+k_y+\frac{e}{\hbar}A_y(x)\right)
\eeq
which fulfill the commutation relation $\left[ a, a^\da\right]=1.$ In the forthcoming analysis, we solve the above
equation each region.

In region $3$ ($d_2 < x < d_3)$, the vector potential $A_y(x)$ is
given by $\frac{\hbar}{e l_{B}^2}x$.
Using the envelope function
$\psi_{B_1}(x,k_y)\equiv\psi_{B_1}(X)$,
where $X=\frac{x}{l_{B}}+k_yl_B$,  \eqref{eqgnrl}
becomes
\begin{equation}\label{eqslt}
\left[-\partial_{X}^2+X^2-1-2\lambda_+\right]\left[-\partial_{X}^2+X^2-1-2\lambda_-\right]\psi_{B_1}(X)=0
\end{equation}
where we have set
\begin{equation}
\lambda_{\tau}=-\frac{1}{2}+\frac{(E-V_{3})^2+\delta_{3}^2}{2\vartheta_{0}^2}+\tau
\frac{\sqrt{(\vartheta_{0}^2-2(E-V_{3})\delta_{3})^2+\gamma_{1}^2
((E-V_{3})^2-\delta_{3}^2)}}{2\vartheta_{0}^2}.
\end{equation}
We solve \eqref{eqslt} 
to obtain the energy in
this region as
\begin{eqnarray}\label{e1}
E&=&V_{3}+\frac{1}{\sqrt{6}}\left[\pm\left[\mu^{\frac{1}{3}}+(A^2+3C)\mu^{\frac{-1}{3}}+2A\right]^\frac{1}{2}\right.\\
&&\left.\pm\left[-6B\sqrt{6}
\left(\mu^{\frac{1}{3}}+(A^2+3C)\mu^{\frac{-1}{3}}+2A\right)^\frac{-1}{2}-\left(\mu^{\frac{1}{3}}+(A^2+3C)\mu^{\frac{-1}{3}}-4A\right)\right]^\frac{1}{2}
\right]\nonumber
\end{eqnarray}
where we have defined the following quantities
\begin{eqnarray}
&&
{\mu=-A^3+27B^2+9AC+3\sqrt{3}\left[-\left(A^2+3C\right)^3+\left(-A^3+27B^2+9AC\right)^2\right]^\frac{1}{2}}\\
&&{A=\delta_{3}^2+(2n+1)\vartheta_{0}^2+\frac{\gamma_{1}^2}{2}}\\
&& {B=\vartheta_{0}^2\delta_{3}}\\
&& {C=\left((2n+1)\vartheta_{0}^2-\delta_{3}^2\right)^2-
\vartheta_{0}^4+\gamma_{1}^2\delta_{3}^2}
\end{eqnarray}
$n$ is an integer number, with $n=\lambda_\tau$.
It is important to note that when $\gamma_1\longrightarrow 0$, the
energy will be reduced to the case of a monolayer graphene.

In region $j=1, 2, 4, 5$, the associated vector potential
$A_y(x)$ is constant and equal to $\frac{\hbar}{el_{B}^2}d$ with
\begin{equation} \label{eq d}
d=\left\{\begin{array}{lll} {d_{2}}, & \mbox{if} & {x < d_{2}} \\
{d_{3}}, & \mbox{if} & {x > d_{3} }. \\
\end{array}\right.
\end{equation}
The corresponding energy
can be written as
\begin{equation}
E-V_{j}= \pm\sqrt{\left(\hbar v_F
k_{j}\right)^2+\frac{\gamma_{1}^2}{2}+\delta_{j}^2\pm\sqrt{\left(\hbar v_F
k_{j}\right)^2(\gamma_{1}^2+4\delta_{j}^2)+\frac{\gamma_{1}^4}{4}}}
\end{equation}
and 
the wave vector $k_j$ is
\beq
k_{j}=\sqrt{\left(\alpha_{j}^\pm\right)^2+\left(k_y+\frac{d}{{l_B}^2}\right)^2}.
\eeq
In the incident region $\alpha_{1}^\pm$ being the wave vector of
the propagating wave, where there are two right-going propagating
modes and two left-going propagating modes. For the transmission
region, $\alpha_{5}^\pm$ is the wave vector of the propagating
wave with two right-going propagating modes.

\section{Transmission probability}

Next we will calculate the transmission probability of electrons
across the double potential barrier in our AB-stacked  bilayer graphene
system. In doing so, we follow two steps where firstly we write our obtained
eigenspinors in  matrix notation and secondly we impose the continuity
of the wave function at each
potential interface.


Recall that our eigenspinors can be obtained in similar way as we have done
in \cite{Jellal} dealing with the same system but scattered by single barrier potential.
To go further, it is convenient to use the matrix formalism such that
the wave function in each region, denoted by the integer $j$,
 can then
be writhen as 
\begin{equation}\lb{psimn}
\psi_j=G_j \cdot M_j\cdot A_j
\end{equation}
where  $A_{1}$ and
$A_{5}$ 
\begin{equation}
 A_{1}^\tau=\left(%
\begin{array}{c}
  \delta_{\tau,1} \\
  r_{+}^\tau \\
  \delta_{\tau,-1} \\
  r_{-}^\tau \\
\end{array}%
\right),\qquad A_{5}^\tau=\left(%
\begin{array}{c}
  t_{+}^\tau \\
  0 \\
  t_{-}^\tau \\
  0 \\
\end{array}%
\right)
\end{equation}
are given in terms of the transmission  $t_{\pm}^\tau$ and reflection  $r_{\pm}^\tau$ amplitudes
as well as the   Kronecker delta symbol $\delta_{\tau,\pm1}$.
For the remanning regions, we have
\beq
A_{2, 3, 4}=\left(\alpha_{2, 3, 4},\alpha_{2, 3,
4}^{'},\beta_{2, 3, 4},\beta_{2, 3, 4}^{'}\right)^T
\eeq
where the appearing coefficients are coming from the decomposition of the eigenspinors in different regions.
In regions
$j=1,\ 2,\ 4,\ 5$, $G_{j}$ and $M_{j}$ take the forms
\begin{equation}
G_{j}=\left(%
\begin{array}{cccc}
  f_{j}^{++} & f_{j}^{+-} & f_{j}^{-+} & f_{j}^{--} \\
  1 & 1  & 1 & 1 \\
 h_j^{+} &  h_j^{+}  &  h_j^{-} &  h_j^{-}  \\
 g_{j}^{+-} & g_{j}^{++} & g_{j}^{--} & g_{j}^{-+}   \\
\end{array}%
\right) ,\qquad
M_{j}=\left(%
\begin{array}{cccc}
  e^{i\alpha_{j}^{+}x} & 0 & 0 & 0 \\
  0 &  e^{-i\alpha_{j}^{+}x} & 0 & 0 \\
  0 & 0 &  e^{i\alpha_{j}^{-}x} & 0 \\
  0 & 0 &  0 &   e^{-i\alpha_{j}^{-}x} \\
\end{array}%
\right)
\end{equation}
with the quantities
\bqr
&&f_{j}^{\tau\pm}=\hbar
v_F\left(\pm\alpha_{j}^{\tau}-i\left(k_y+\frac{d}{l_{B}^2}\right)\right)/\left(E-V_j-\delta_j\right)\\
&& h_{j}^\tau=\frac{E-V_j-\delta_j}{\gamma_1}\left[1-\frac{\left(\hbar
v_F\right)^2\left[\left(\alpha_{j}^{\tau}\right)^2+\left(k_y+\frac{d}{l_{B}^2}\right)^2\right]^2}{\left(E-V_j-\delta_j\right)^2}\right]\\
&& g_{j}^{\tau\pm}=-\frac{E-V_j-\delta_j}{E-V_j+\delta_j}h_{j}^\tau
f_{j}^{\tau\pm}.
\eqr
In region $j=3$,
we have $G_{3}=\mathbb{I}_{4}$ and $M_{3}$ reads as
\begin{equation}
M_{3}=\left(%
\begin{array}{cccc}
\eta_-\lambda_+\chi_{-1}^{++} &\eta_{-}^*\lambda_+\chi_{-1}^{+-}&
\eta_-\lambda_-\chi_{-1}^{-+}&\eta_{-}^*\lambda_-\chi_{-1}^{--}\\
\chi_{0}^{++} &\chi_{0}^{+-}& \chi_{0}^{-+} &\chi_{0}^{--} \\
\zeta^+\chi_{0}^{++} &\zeta^+\chi_{0}^{+-}& \zeta^-\chi_{0}^{-+} &\zeta^-\chi_{0}^{--} \\
\eta_{+}^*\zeta^+\chi_{1}^{++} &\eta_{+}\zeta^+\chi_{1}^{+-}&
\eta_{+}^*\zeta^-\chi_{1}^{-+} &\eta_{+}\zeta^-\chi_{1}^{--}
\end{array}
\right)
\end{equation}
where we have set
\bqr
&& \eta_{\pm}=\frac{-i\sqrt{2}\vartheta_0}{E-V_{3}\pm\delta_{3}}\\
&&\zeta^\tau=\frac{E-V_{3}-\delta_{3}}{\gamma_1}-\frac{2\vartheta_{0}^2\lambda_\tau}{\gamma_1\left(E-V_{3}-\delta_{3}\right)}\\
&& \chi_{l}^{\tau\pm}=D\left[\lambda_\tau+l,\pm \sqrt{2}X\right]
\eqr
 $D\left[\lambda_\tau+l,\pm \sqrt{2}X\right]$ are the parabolic cylindrical functions with the quantum numbers
 $\lambda_\tau=n$ and $l=0,\pm 1.$

Now let us
consider the
 boundary conditions at $x = d_j$ and then from \eqref{psimn} we end up with the set
 of equations 
\begin{eqnarray}
&&{G_1\cdot M_1[x=d_1]\cdot A_{1}^\tau=G_{2}\cdot M_{2}[x=d_1]\cdot
A_{2}}\\
&&{G_{2}\cdot M_{2}[x=d_2]\cdot A_{2}=G_{3}\cdot
M_{3}[x=d_2]\cdot A_{3}}\\
&&{G_{3}\cdot M_{3}[x=d_3]\cdot
A_{3}=G_{4}\cdot M_{4}[x=d_3]\cdot A_{4}}\\
&&{G_{4}\cdot
M_{4}[x=d_4]\cdot A_{4}=G_{5}\cdot M_{5}[x=d_4]\cdot A_{5}^\tau}.
\end{eqnarray}
Using the transfer matrix method we can connect $A_{1}^\tau$ with
$A_{5}^\tau$ through the matrix $N$
\begin{equation}
N=\prod_{j=1}^{4} M_{j}^{-1}[x=d_{j}]\cdot G_{j}^{-1}\cdot
G_{j+1}\cdot M_{j+1}[x=d_{j}]
\end{equation}
which can help to
explicitly determine $t_{\pm}^\tau$ and therefore the  corresponding transmission probability, in a compact form, as
\begin{equation}
T_{\pm}^\tau
=\frac{\alpha_{2}^\pm}{\alpha_{1}^\tau}\mid
 t_{\pm}^\tau\mid^2.
\end{equation}

In next section, we will study numerically two interesting cases depending on the
value of the incident energies, $E$, as compared with the
interlayer coupling parameter $\gamma_1$. The two band tunneling
leads to one transmission and one reflection channel and takes
place at energies less than the interlayer coupling, $E <
\gamma_1$. On the other hand, for energies higher than the
interlayer coupling parameter $\gamma_1$, $E > \gamma_1$, the four
band tunneling takes place giving rise to four transmission and
four reflection channels.

\section{Numerical results}

\begin{figure}[h!]
 \centering
\includegraphics[width=6cm, height=5cm]{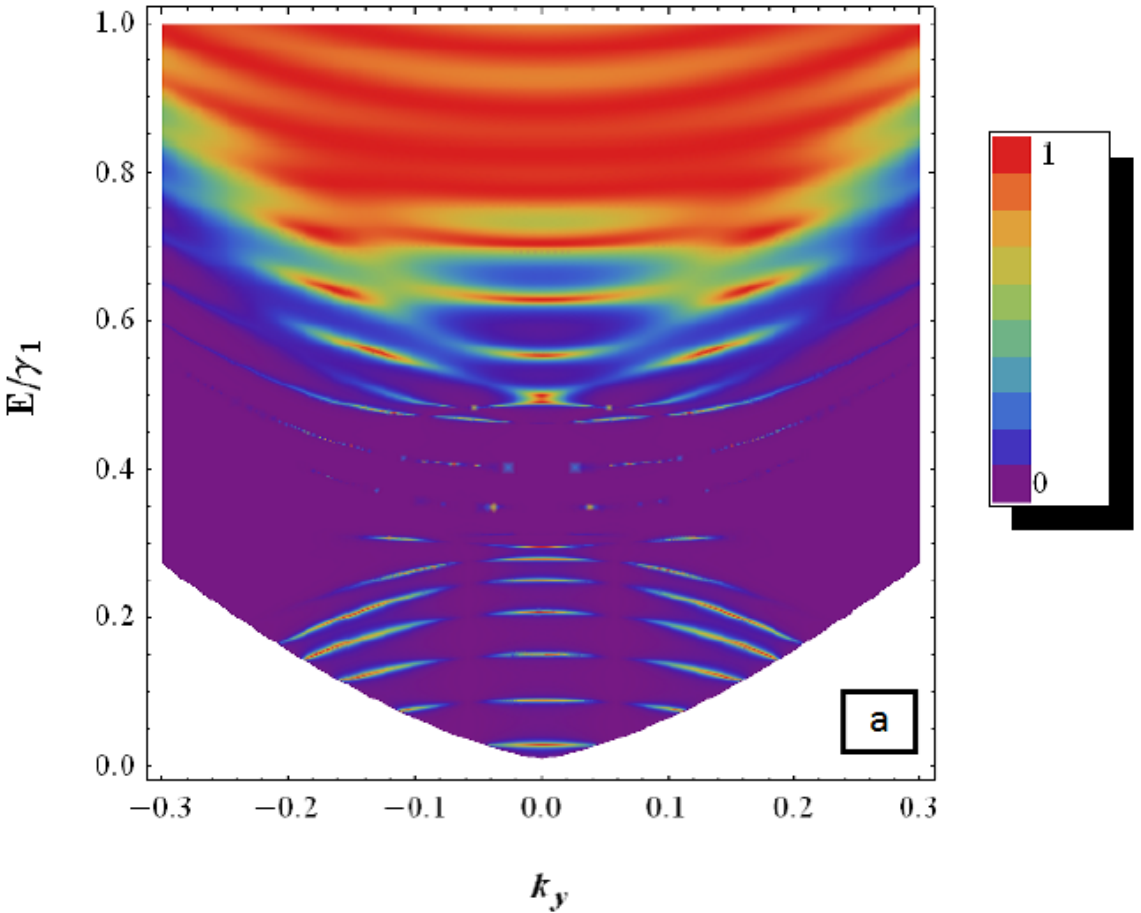}
 ~~~~~~ \includegraphics[width=6cm, height=5cm]{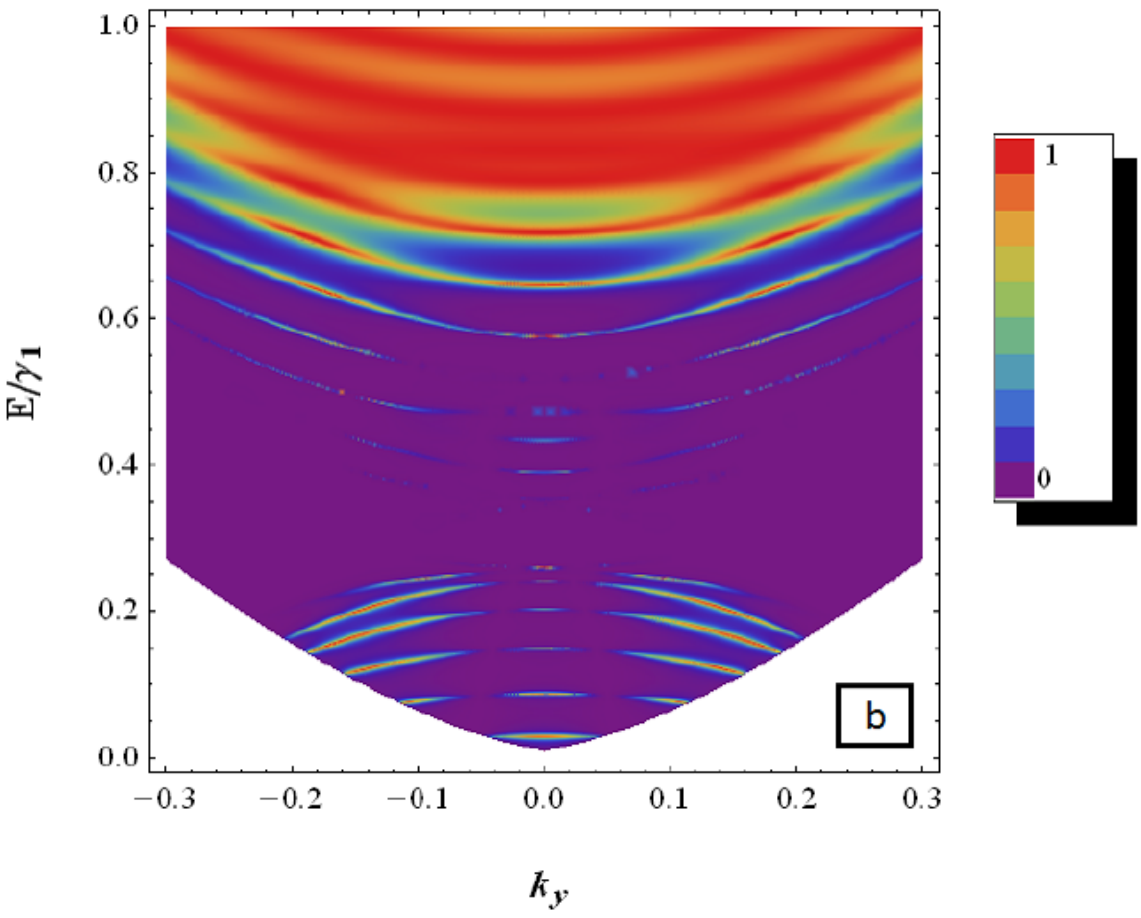}
\\ \includegraphics[width=6cm, height=5cm]{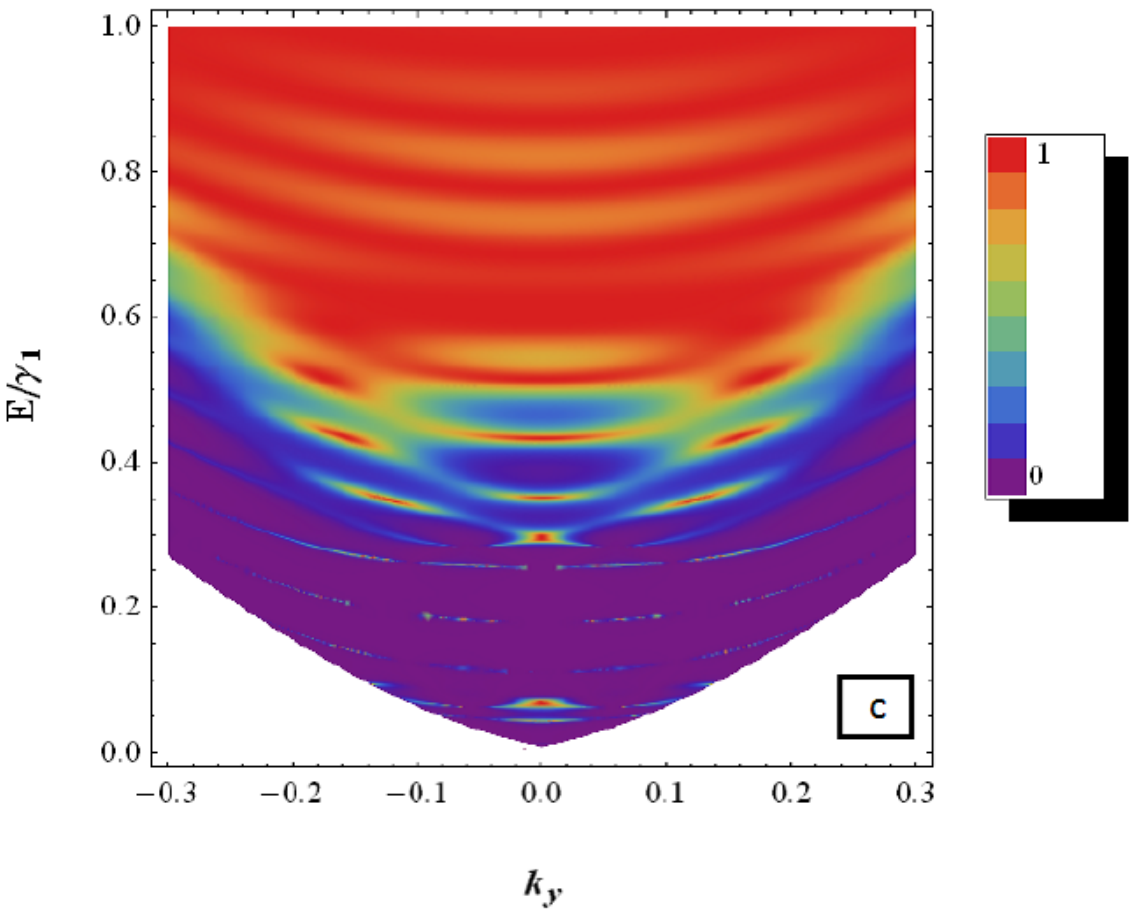}
 ~~~~~~ \includegraphics[width=6cm, height=5cm]{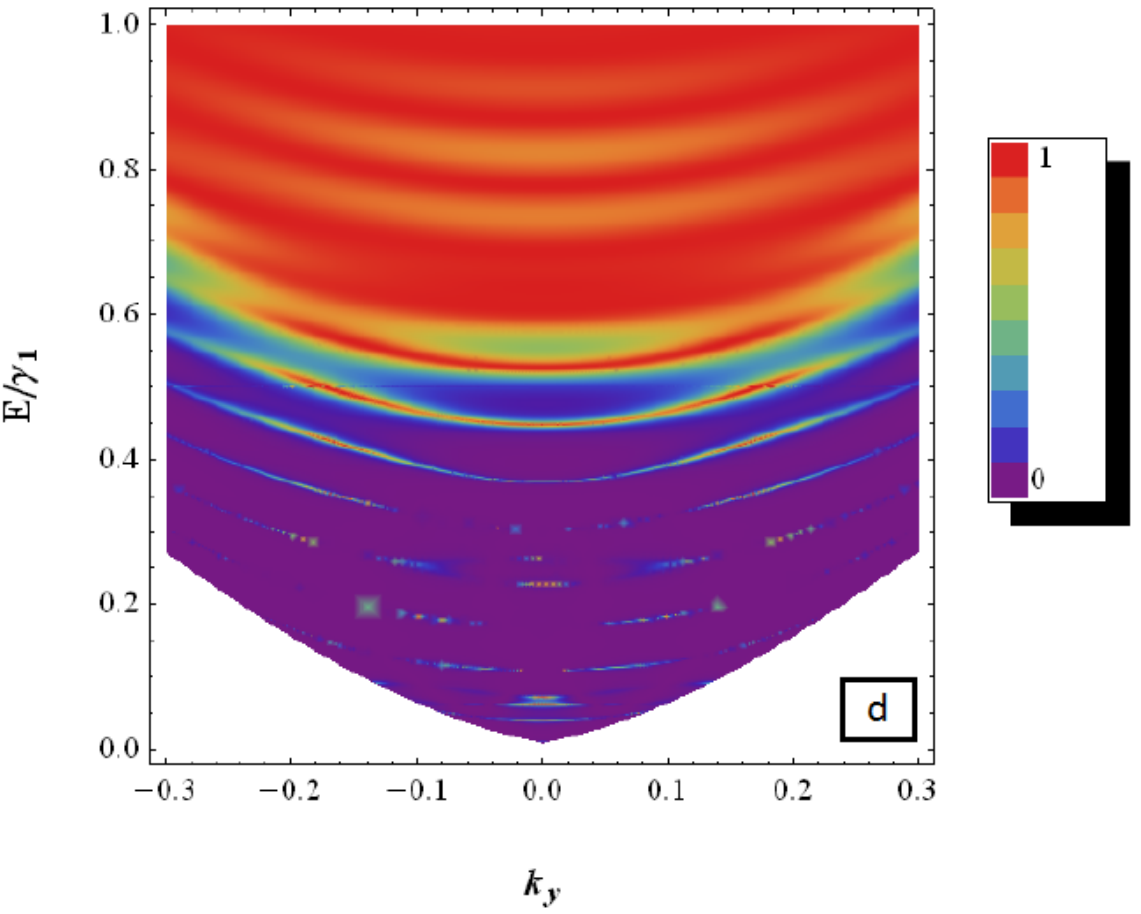}
\caption{Density plot of transmission probability as a function of
the transverse wave vector $k_y$ and its energy $E$, for
$d_4=-d_1= 30\ nm $, $d_3=-d_2=20\ nm$
and $l_B = 18.5\ nm$. (a): $V_2 = V_4 = 0.6\ \gamma_1$, $V_3 =
0.3\ \gamma_1$ and $\delta_2 = \delta_3 = \delta_4 = 0\ \gamma_1$.
(b): the same parameters as in (a) with $\delta_2 = \delta_4 =
0.1\ \gamma_1$ and $\delta_3 =0.05\ \gamma_1$. (c): $V_2 = V_4 =
0.4\ \gamma_1$ and $V_3=\delta_2 = \delta_3 = \delta_4 =0\
\gamma_1$. (d): the same parameters as in (c) but with
$\delta_2=\delta_4=0.1\ \gamma_1$.} \label{fig.TB3}
\end{figure}

We compute numerically the transmission probability through the
double barrier in the presence of a magnetic field in the low
energy regime. The two band model only allows for one mode of
propagation, leading to one transmission. In this case, Figure
\ref{fig.TB3} shows the density plot of the transmission
probability as a function of the transverse wave vector $k_y$ and
energy $E$.
Figures (\ref{fig.TB3}a) and (\ref{fig.TB3}c) present two different structures
of the double barrier.
In Figure  (\ref{fig.TB3}a), we show that the transmission for
$V_2 = V_4 = 0.6\ \gamma_1$, $V_3 = 0.3\ \gamma_1$ and $\delta_2 =
\delta_3 = \delta_4 = 0\ \gamma_1$. At nearly normal incidence ($
k_y \approx - \frac{d_2}{l_{B}^2}$ and $ k_y \approx -
\frac{d_3}{l_{B}^2}$) the transmission is zero and there are no
resonances in the low regime energy  $E < V_3$. While resonances
occur at non-normal incidence, which is equivalent to the case
of a single barrier \cite{Jellal}. In the regime of energy $V_3 < E
< V_2 = V_4$, it is clearly seen that the peaks are due to the
bound electron states \cite{Hassan}. In addition, when $E > V_2 =
V_4$ the Dirac fermions exhibit transmission resonances.
While many of these results are similar to those obtained in
Figure  (\ref{fig.TB3}c), with $V_3=\delta_3=0\ \gamma_1$. It is
important to see how the interlayer potential difference and the resulting
energy gap in the energy spectrum affect the
transmission probability. 
To answer this inquiry
we give the results
presented in Figures  (\ref{fig.TB3}b) and (\ref{fig.TB3}d).
We notice that in Figure  (\ref{fig.TB3}b), the transmission displays
sharp peaks inside the transmission gap around the Dirac point $E
= V_2 = V_4$, that are absent around $E = V_3$. As observed in
Figure  (\ref{fig.TB3}d) the transmission probability displays
sharp peaks around the only Dirac point at $E = V_2 = V_4$.
It is clearly seen that the displayed sharp peaks inside the transmission
gap for a double barrier structure do not appear in the case for the single barrier
\cite{Jellal}.

\begin{figure}[h!]
 \centering
 \includegraphics[width=6cm, height=5cm]{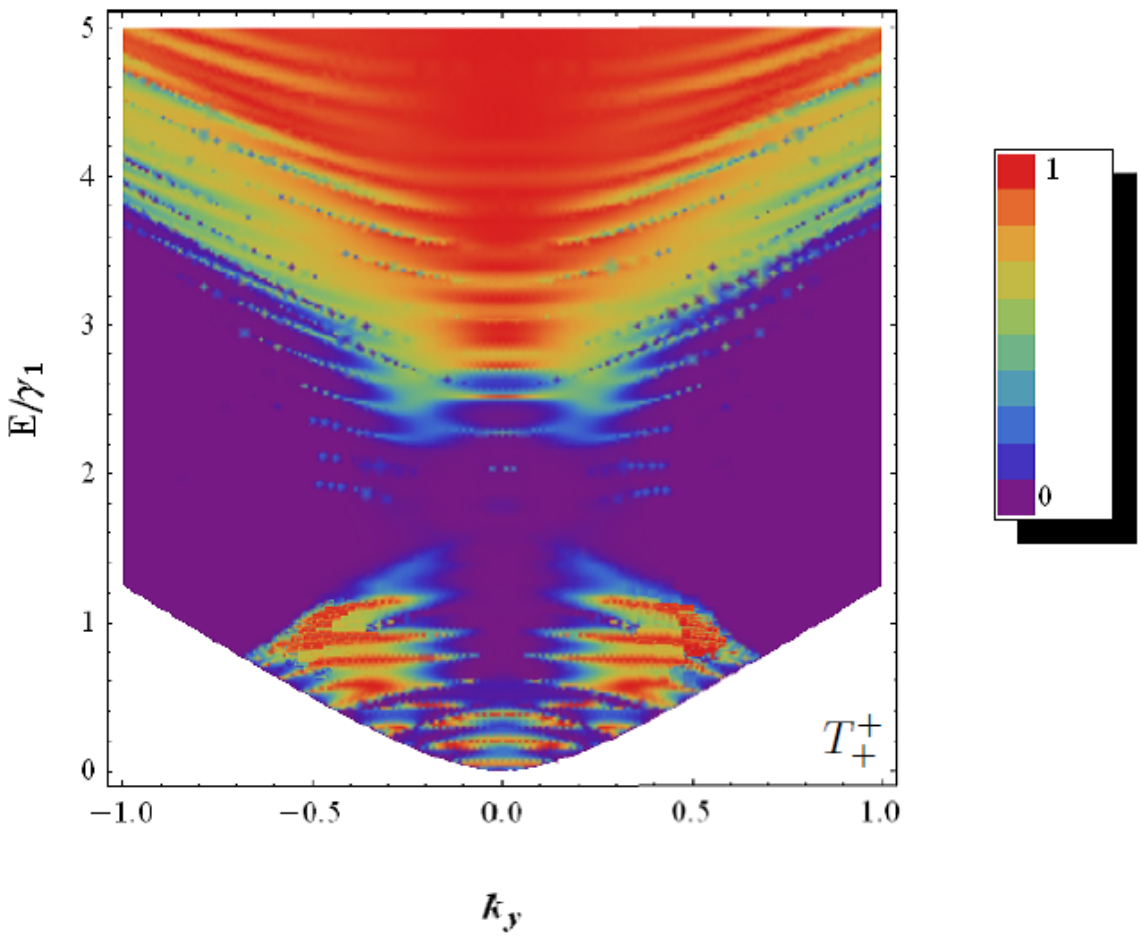}
 ~~~~~~ \includegraphics[width=6cm, height=5cm]{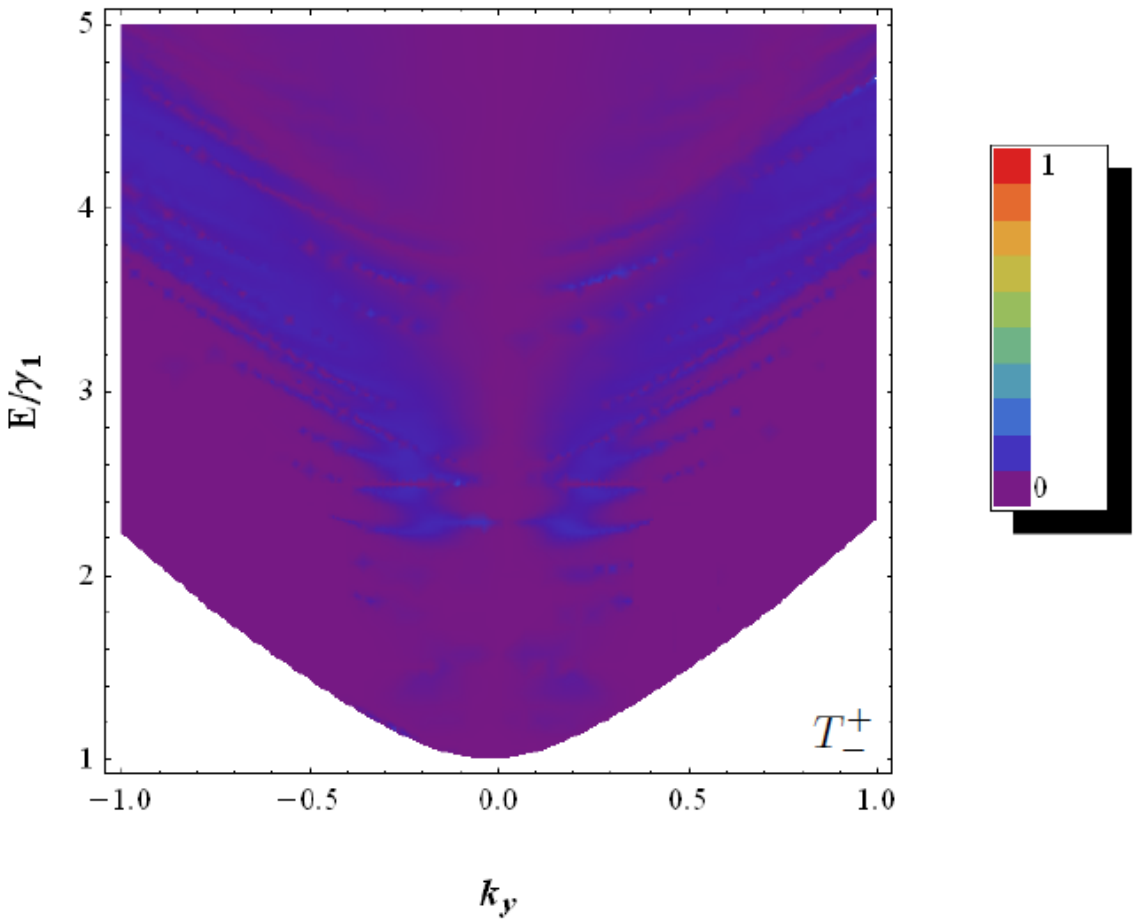}
\\ \includegraphics[width=6cm, height=5cm]{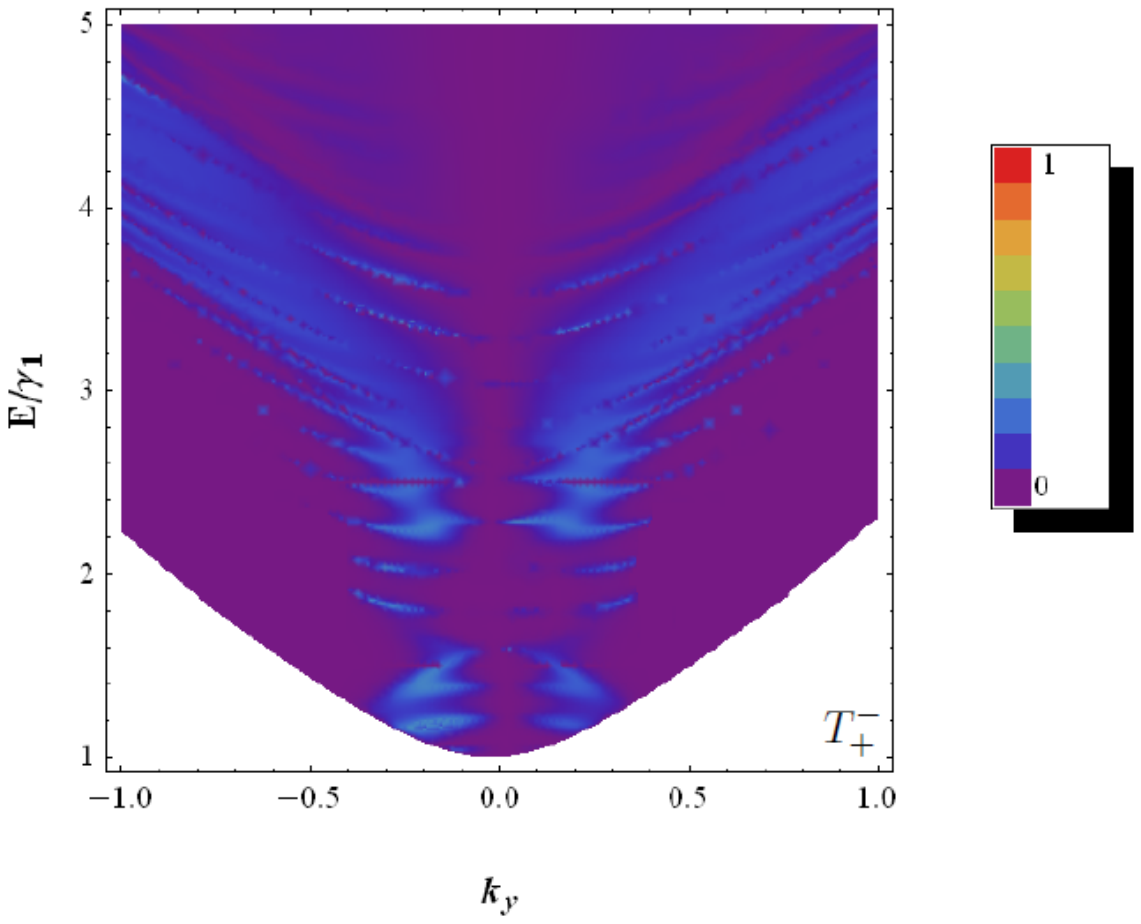}
 ~~~~~~ \includegraphics[width=6cm, height=5cm]{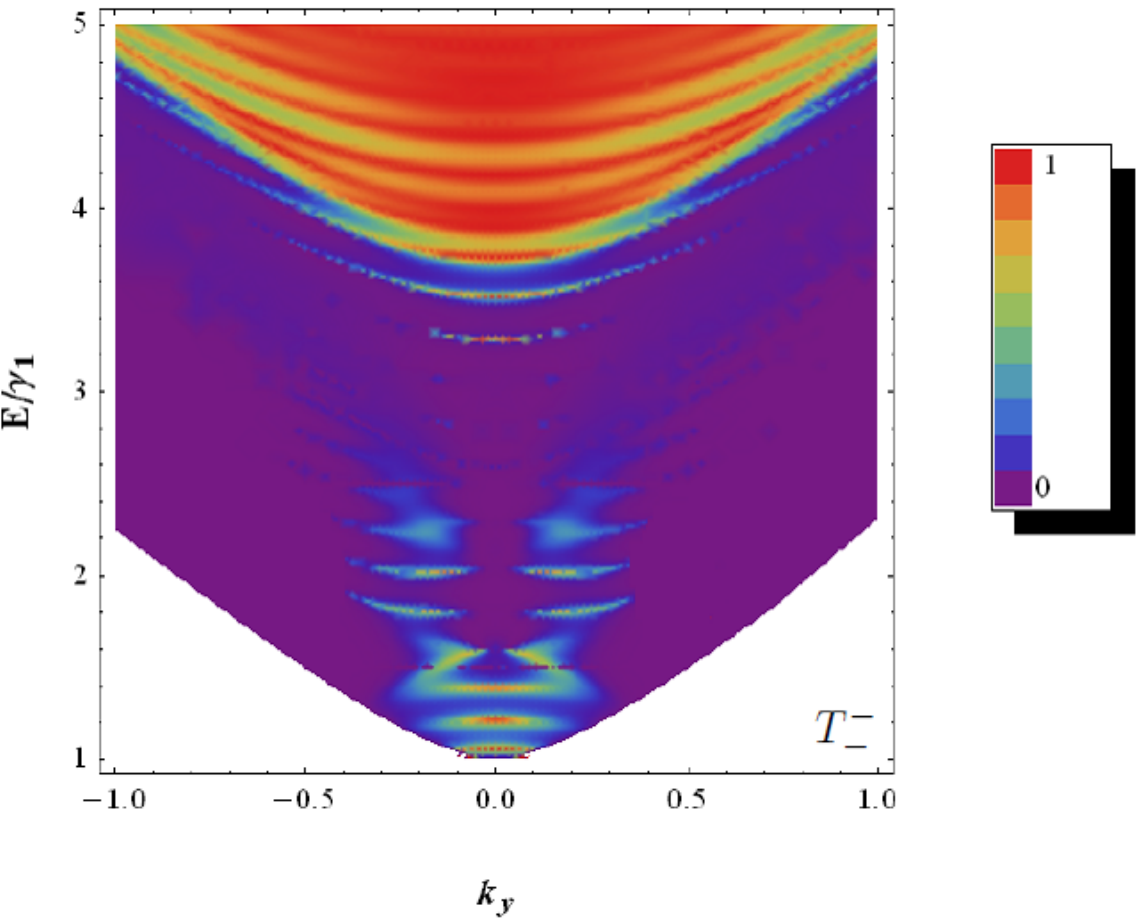}
 \caption{Density plot of the four transmission channels as a function of
the transverse wave vector $k_y$ and its energy $E$, with
$V_2=V_4=2.5\ \gamma_1$, $V_3=1.5\ \gamma_1$,
$\delta_2=\delta_3=\delta_4=0\ \gamma_1$, $l_B=18.5\ nm$,
$d_4=-d_1=14\ nm$ and $d_3=-d_2=7.5\ nm$.} \label{fig.FB3}
\end{figure}

For energies larger than the interlayer coupling, $E > \gamma_1$,
we will have four transmission channels resulting in what we call
the four band tunneling. Therefore, in the transmission channels
$T_{+}^+$ and $T_{-}^- $ electrons propagate via $\alpha^+$ and
$\alpha^-$ mode, respectively. For $T_{-}^+$ scattering from
$\alpha^+$ mode to the $\alpha^-$ mode and $T_{+}^-$ operates in the other
direction \cite{Duppen,Hassan,Jellal}. In Figure \ref{fig.FB3}, we
show the different channels associated with the four transmission
probabilities as a function of the transfer wave vector $k_y$ and
energy $E$. The potential barrier heights are set to be
$V_2=V_4=2.5\ \gamma_1$, $V_3=1.5\ \gamma_1$ and the interlayer
potential difference is zero, i.e. $\delta_2=\delta_3=\delta_4=0\
\gamma_1$. For energies smaller than $V_3-\gamma_1$, there are
propagating $\alpha^+$ states in the region $j=3$, which lead to
a nonzero transmission in the $T_{+}^+$ channel. The cloak effect
\cite{Gu} occurs in the energy region $V_3-\gamma_1 < E < V_3$ for
nearly normal incidence, in $T_{+}^+$ channel $\left(k_y \approx
-\frac{d_3}{l_{B}^2}\ \mbox{and} \ k_y \approx -\frac{d_2}{l_{B}^2}\right)$, in
$T_{-}^+ $ channel $\left(k_y \approx -\frac{d_3}{l_{B}^2}\right)$ and in
$T_{+}^-$ channel $\left(k_y \approx -\frac{d_2}{l_{B}^2}\right)$ where the
two modes $\alpha^+$ and $\alpha^-$ are decoupled and therefore no
scattering occurs between them \cite{Duppen,Hassan}. While for
non-normal incidence the two modes $\alpha^+$ and $\alpha^-$ are
coupled, so that the transmissions $T_{+}^{+}$, $T_{-}^{+}$ and
$T_{+}^{-}$ channels are not zero. The transmission probabilities
$T_{-}^{+}$ and $T_{+}^{-}$ are different ($T_{-}^{+} \neq
T_{+}^{-}$), which introduces the asymmetry of double barrier in
the presence of a magnetic field and $V_3 \neq 0$. In fact, these
transmission probabilities are associated with electrons moving in opposite
direction. For $T_{-}^{-}$ electrons propagate via $\alpha^-$ mode
in the two energy regimes $E < V_3$ and $E > V_2+\gamma_1$, but
is absent in the energy regime $V_2 < E < V_2+\gamma_1$ so that the
transmission is suppressed in this region, which is equivalent
to the cloak effect \cite{Duppen,Hassan,Jellal}. Additionally, in
the four transmission channels the resonances are resulting from the
bound electrons in the well between the two barriers ($V_3 < E <
V_2=V_4$).

\begin{figure}[H]
 \centering
 \includegraphics[width=6cm, height=5cm]{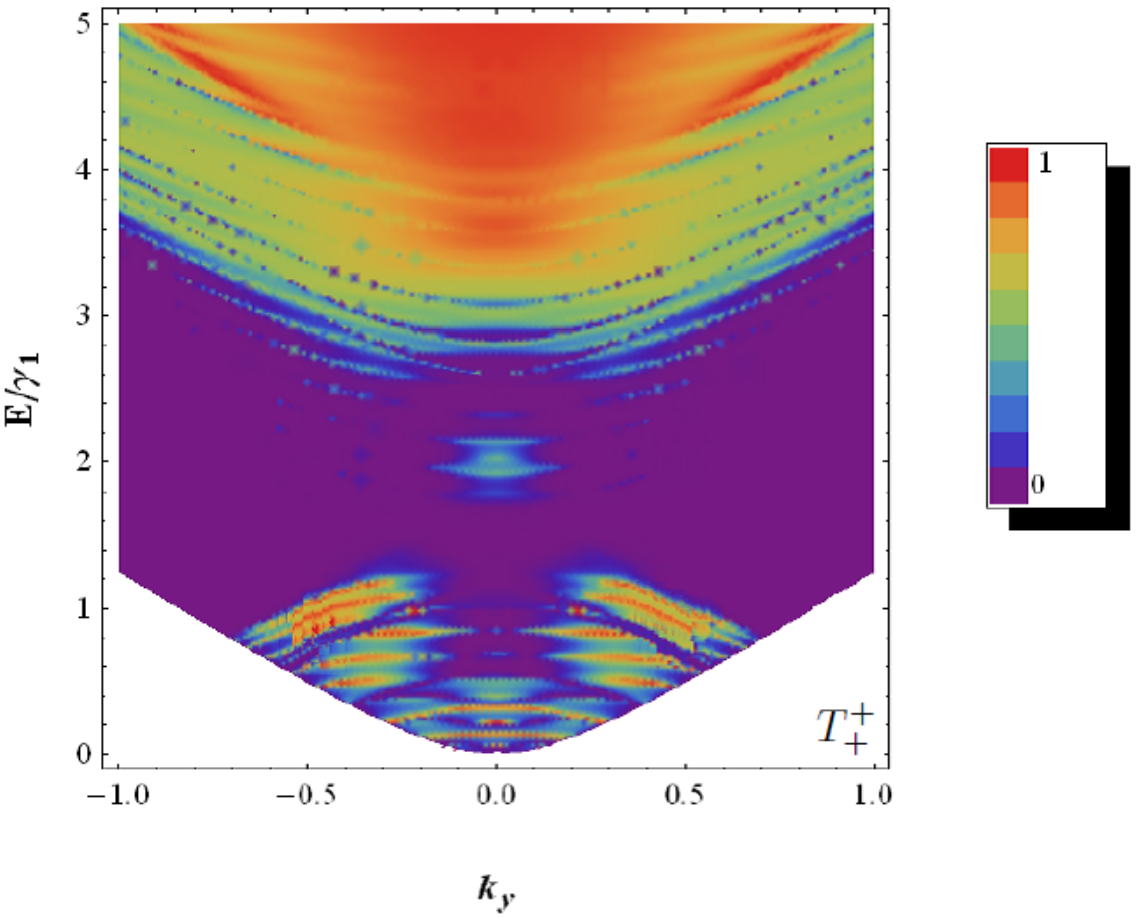}
 ~~~~~~ \includegraphics[width=6cm, height=5cm]{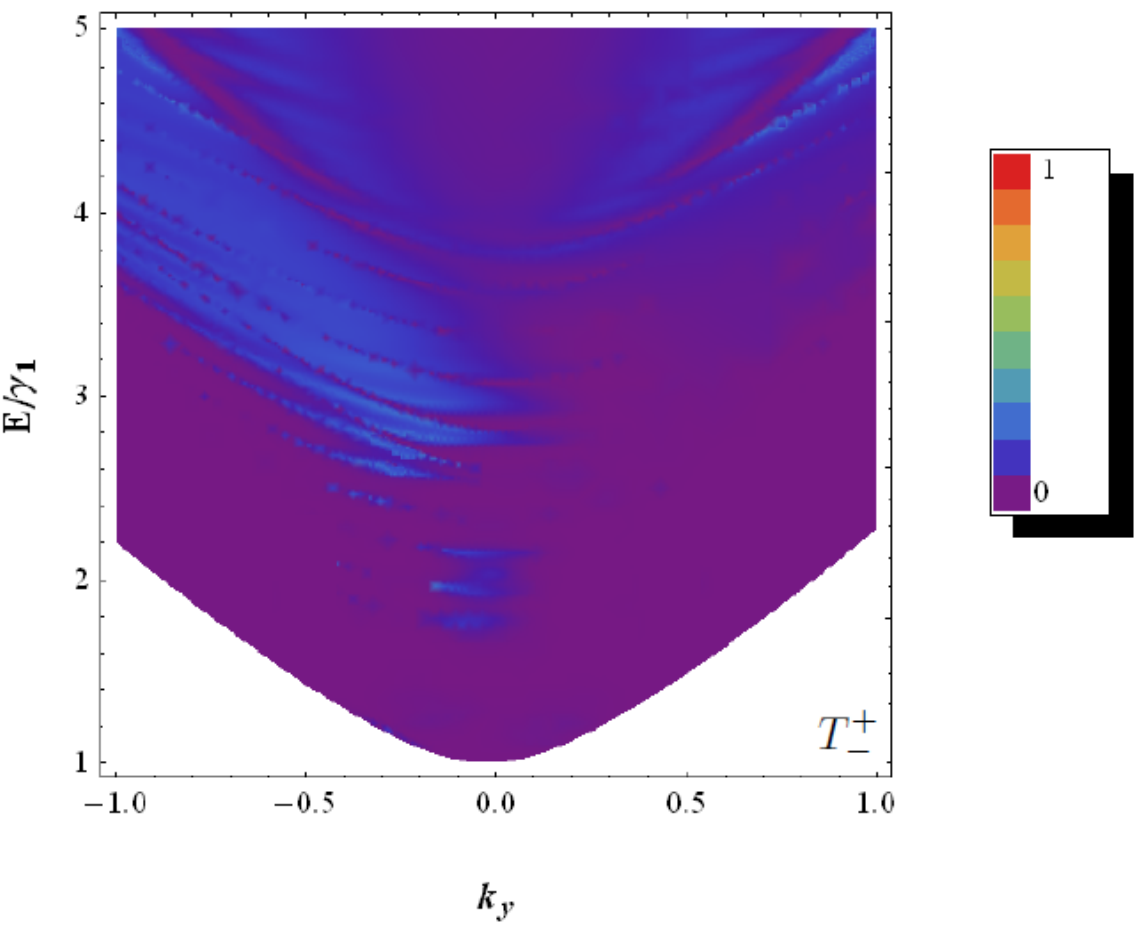}
\\\includegraphics[width=6cm, height=5cm]{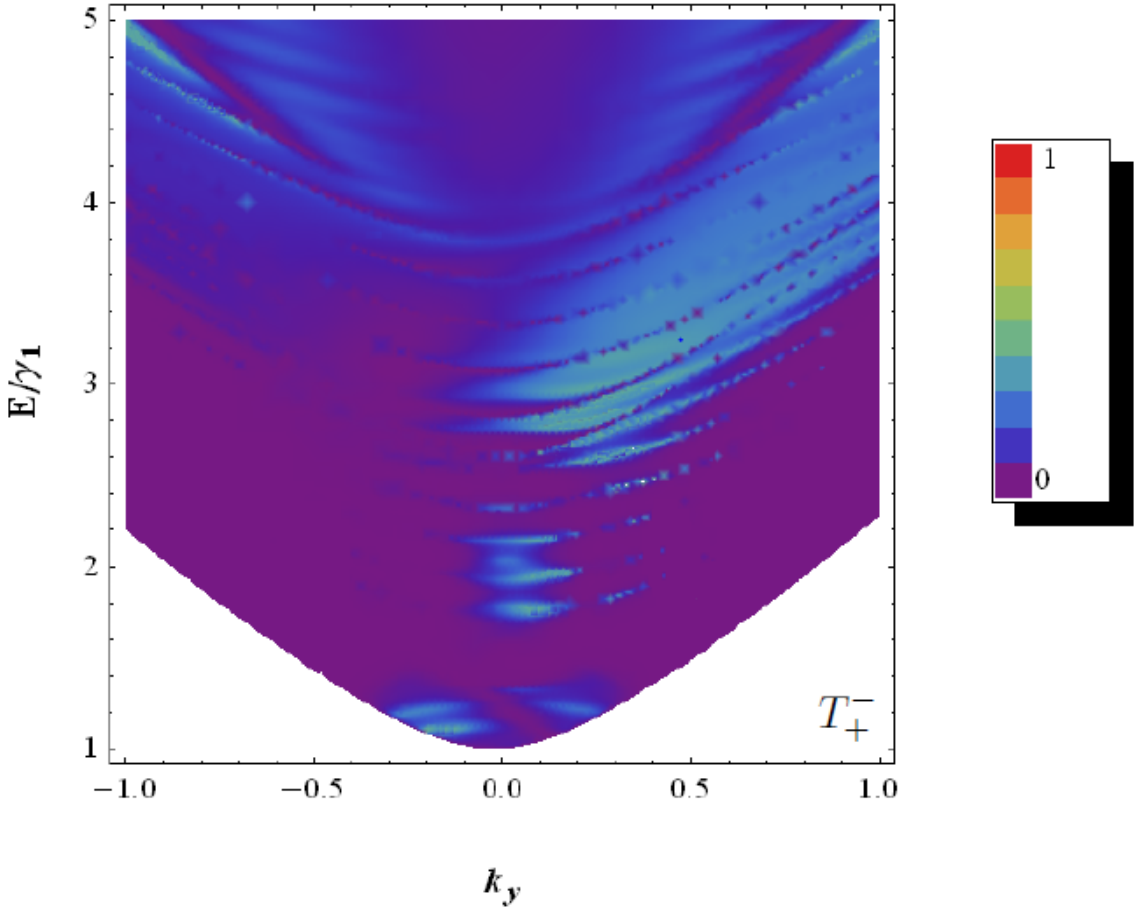}
 ~~~~~~\includegraphics[width=6cm, height=5cm]{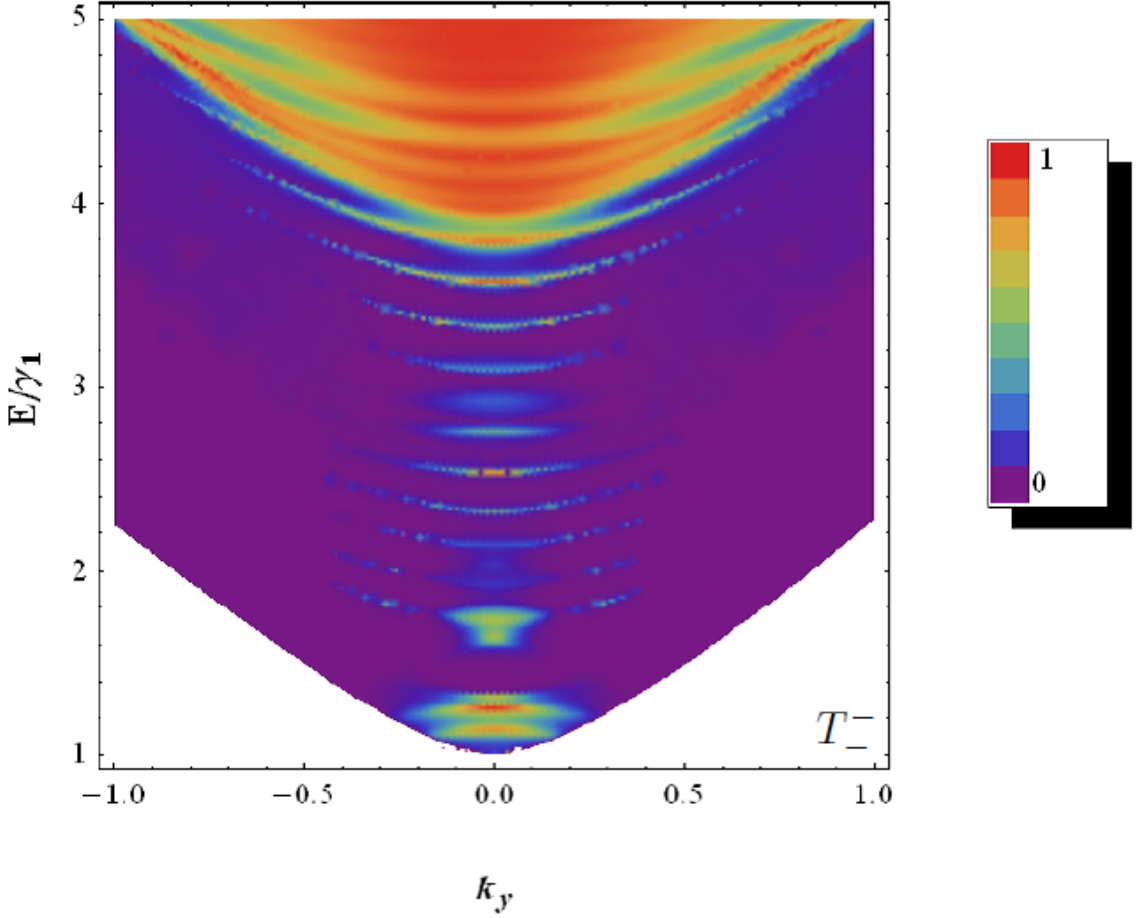}
 \caption{Density plot of the four transmission channels as a function of
the transverse wave vector $k_y$ and its energy $E$, with
$V_2=V_4=2.5\ \gamma_1$, $V_3=1.5\ \gamma_1$,
$\delta_2=\delta_4=0.4\ \gamma_1$, $\delta_3=0.2\ \gamma_1$,
$l_B=18.5\ nm$, $d_4=-d_1=14\ nm$ and $d_3=-d_2=7.5\ nm$.}\label{fig.FB4}
\end{figure}

Now let see how the interlayer potential difference will affect
the different transmission channels. In Figure \ref{fig.FB4},
we present the density plot of the four transmission channels as
function of the transfer wave vector $k_y$ and energy $E$. We consider
the same parameters as in Figure \ref{fig.FB3} but for
$\delta_2=\delta_4=0.4\ \gamma_1$ and $\delta_3=0.2\ \gamma_1$. We
notice that the four transmission channels are related to the
transmission gap around to the Dirac points $E=V_2=V_4$ and $E=V_3$.
It is clearly seen that the transmission display sharp peaks
inside the transmission gap around the Dirac point $E = V_2 =
V_4$. These peaks can be attributed to the bound states formed in
the double barrier structure \cite{Hassan}. That are correlated to
the transmission gap and show a suppression due to cloak effect
around $E = V_3$, as it was the case for the single barrier
\cite{Jellal}.

\begin{figure}[H]
 \centering
 \includegraphics[width=6cm, height=5cm]{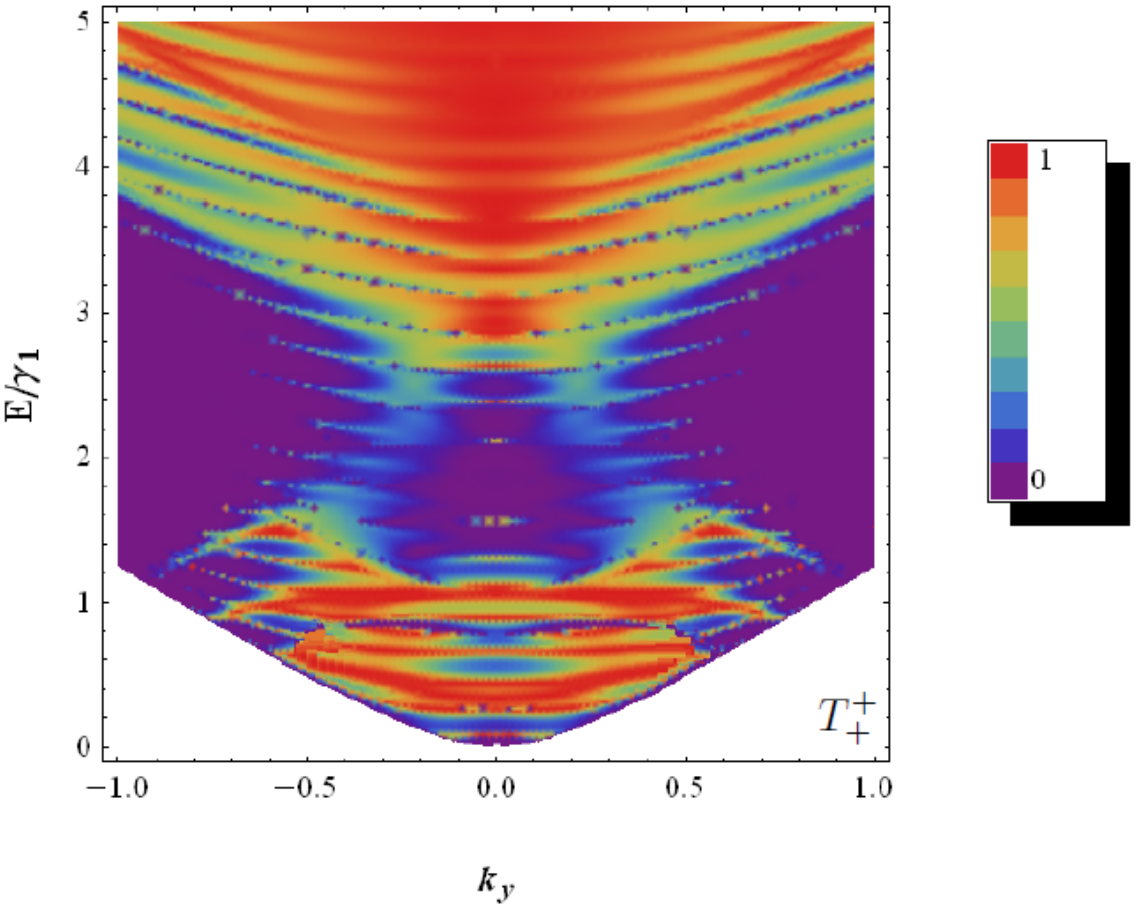}
 ~~~~~~ \includegraphics[width=6cm, height=5cm]{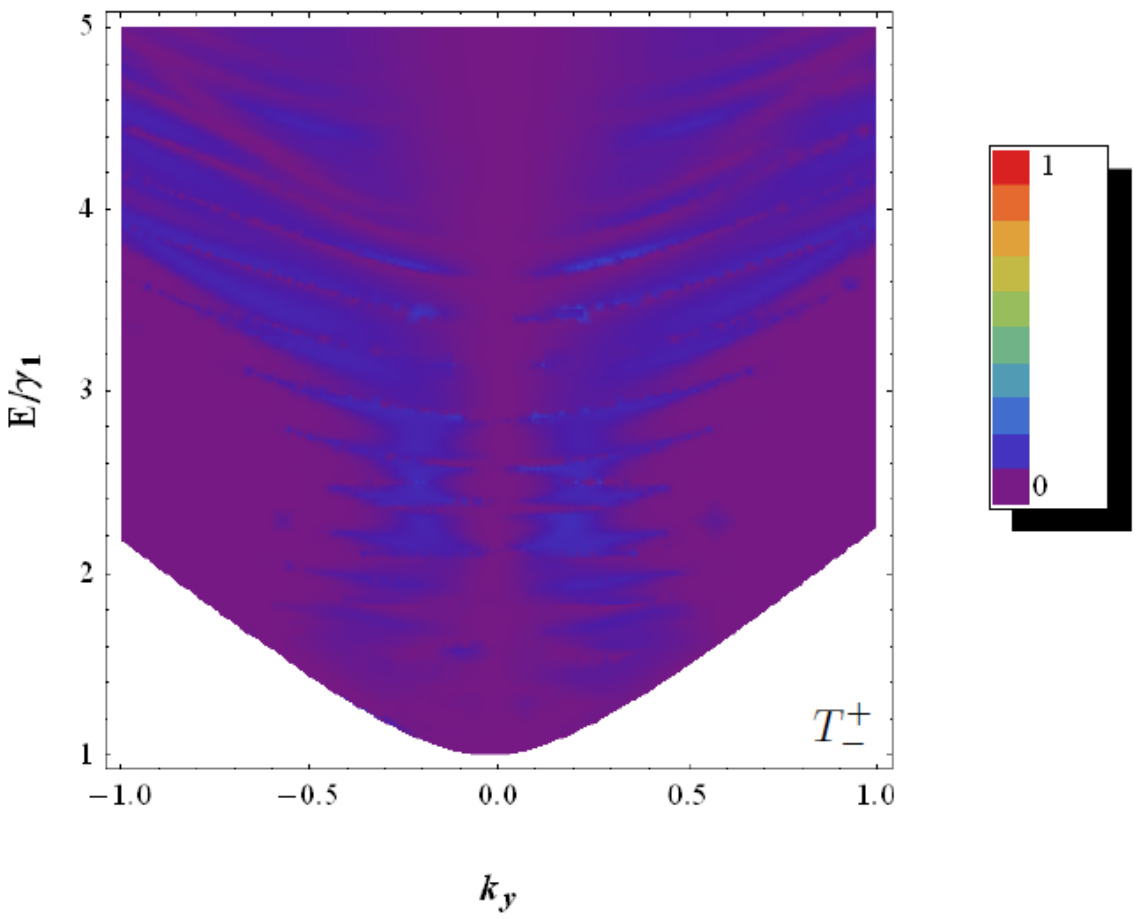}
 \\ \includegraphics[width=6cm, height=5cm]{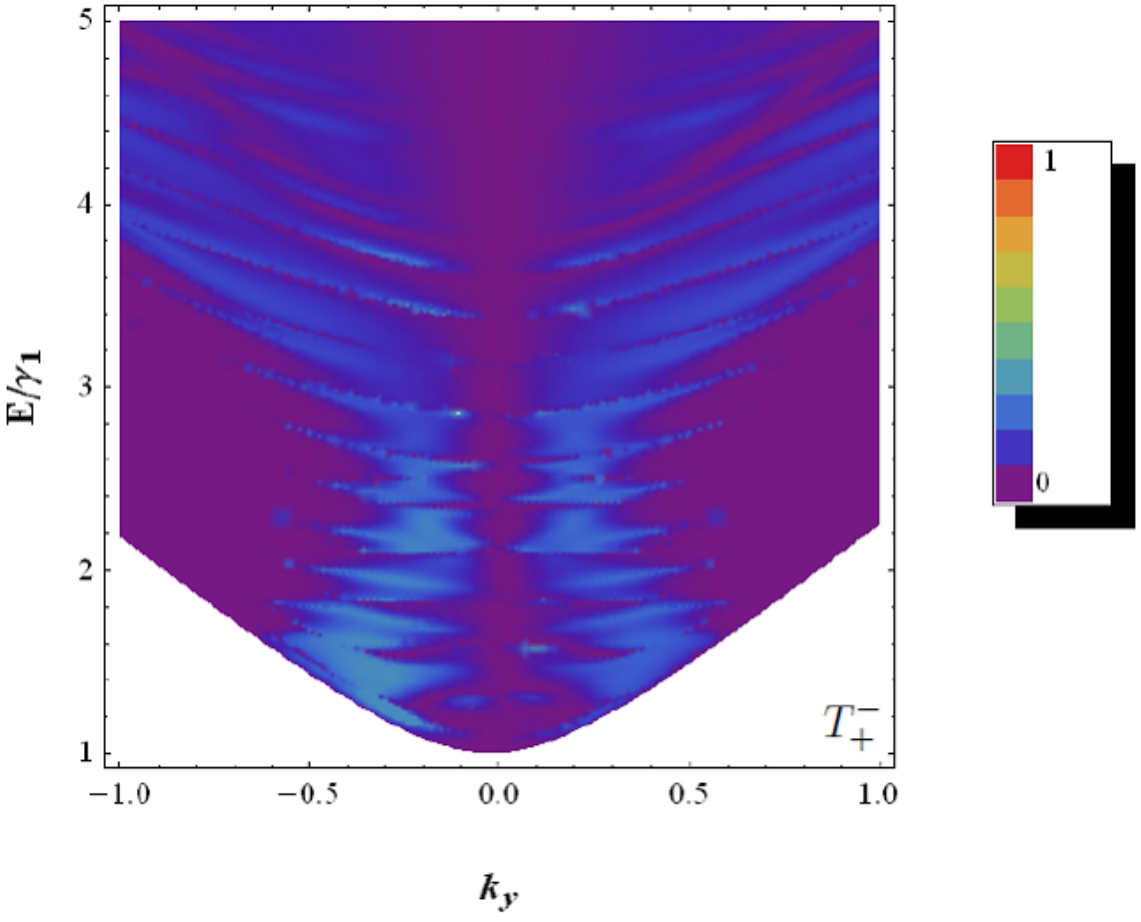}
 ~~~~~~ \includegraphics[width=6cm, height=5cm]{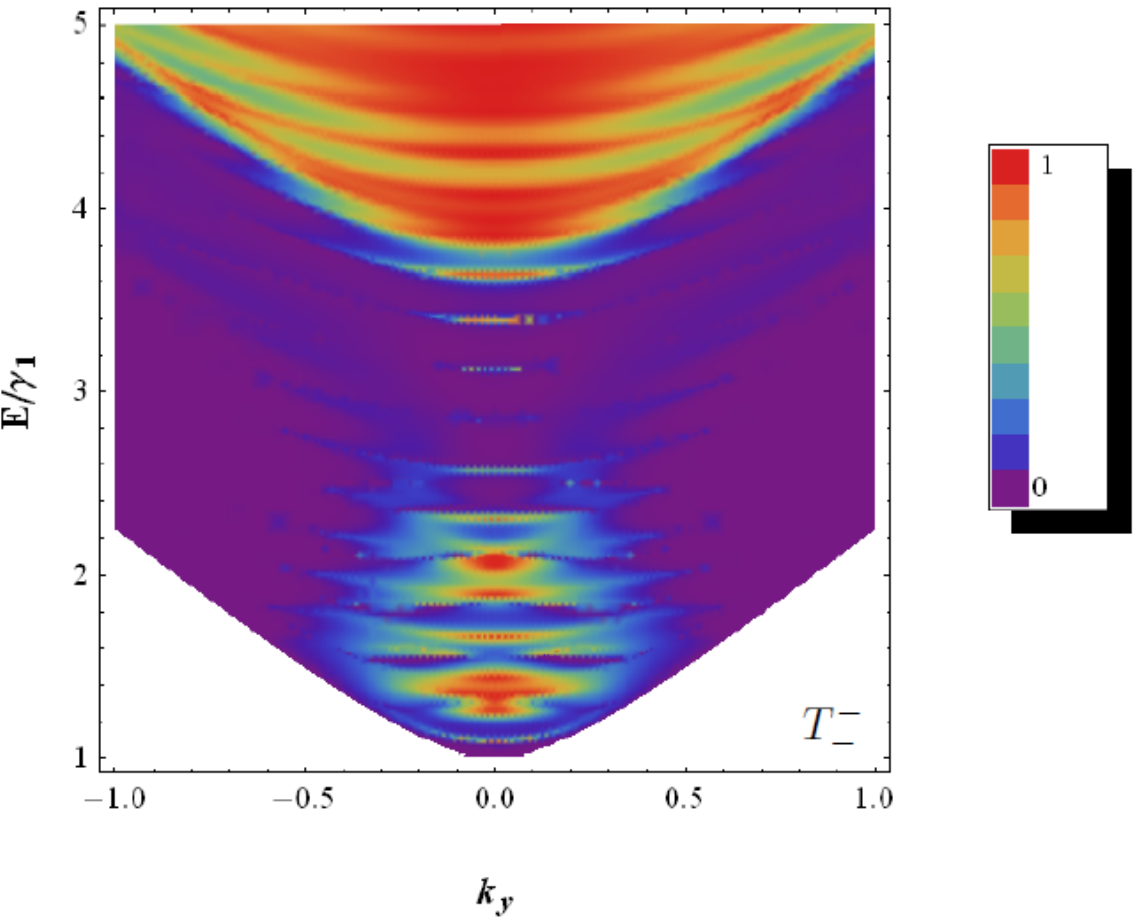}
\caption{Density plot of the four transmission channels as a
function of the transverse wave vector $k_y$ and its energy $E$,
with $V_2=V_4=2.5\ \gamma_1$, $V_3=0\ \gamma_1$,
$\delta_2=\delta_3=\delta_4=0\ \gamma_1$, $l_B=18.5\ nm$,
$d_4=-d_1=14\ nm$ and $d_3=-d_2=7.5\ nm$.} \label{fig.FB1}
\end{figure}

\begin{figure}[h!]
 \centering
 \includegraphics[width=6cm, height=5cm]{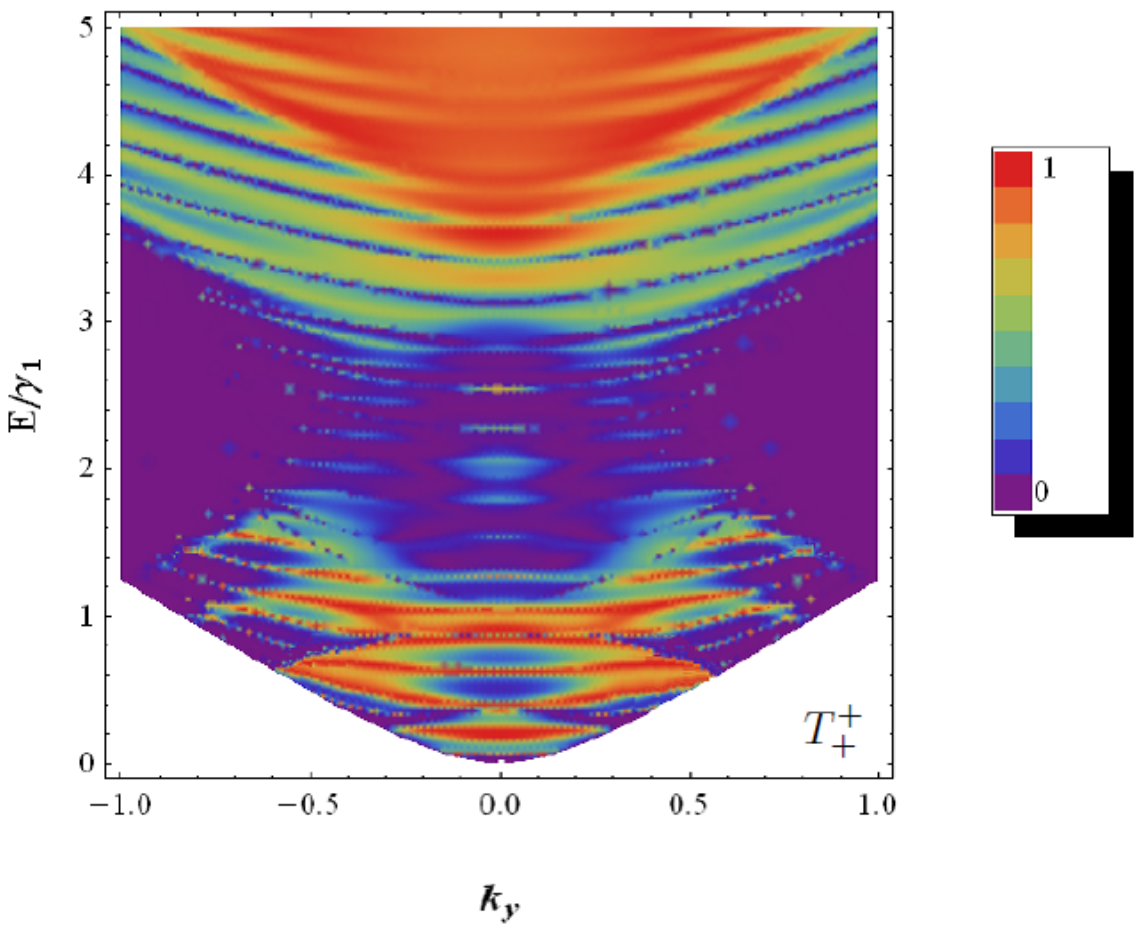}
  ~~~~~~ \includegraphics[width=6cm, height=5cm]{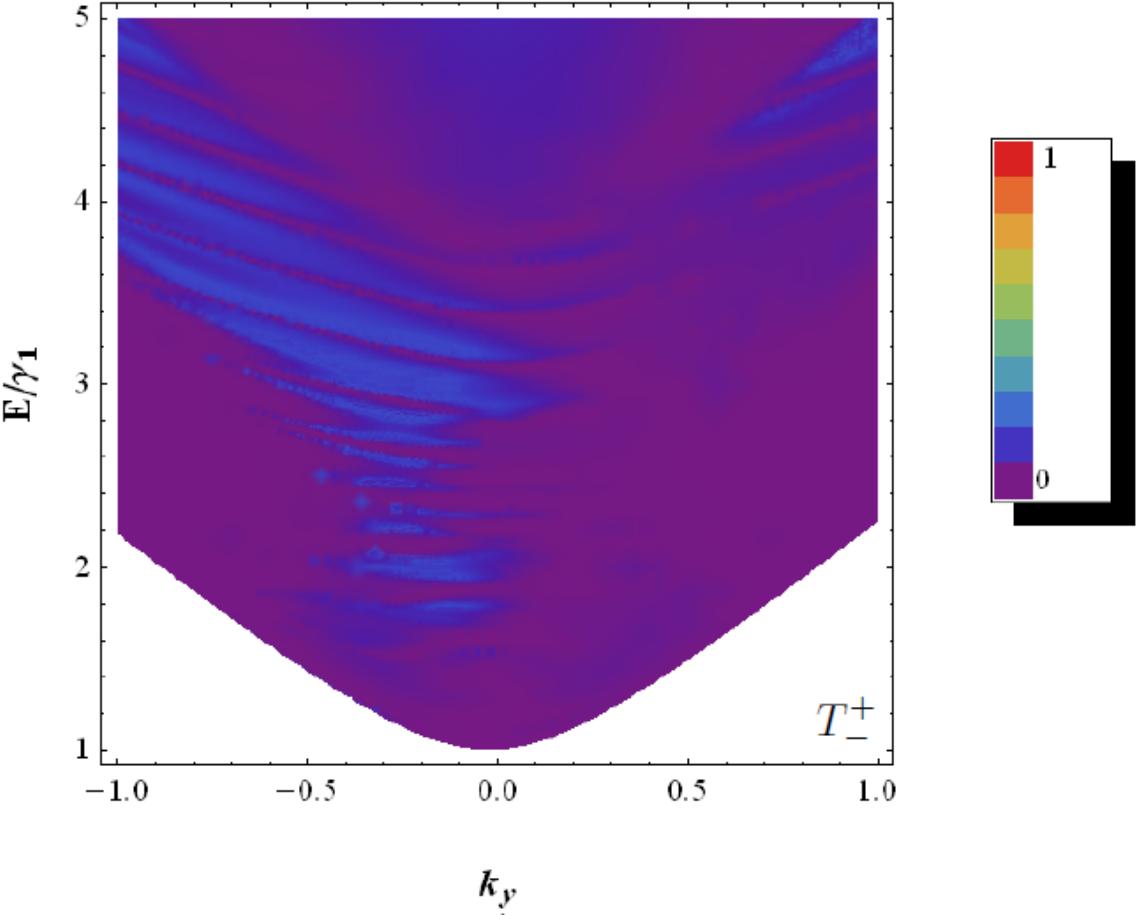}
\\ \includegraphics[width=6cm, height=5cm]{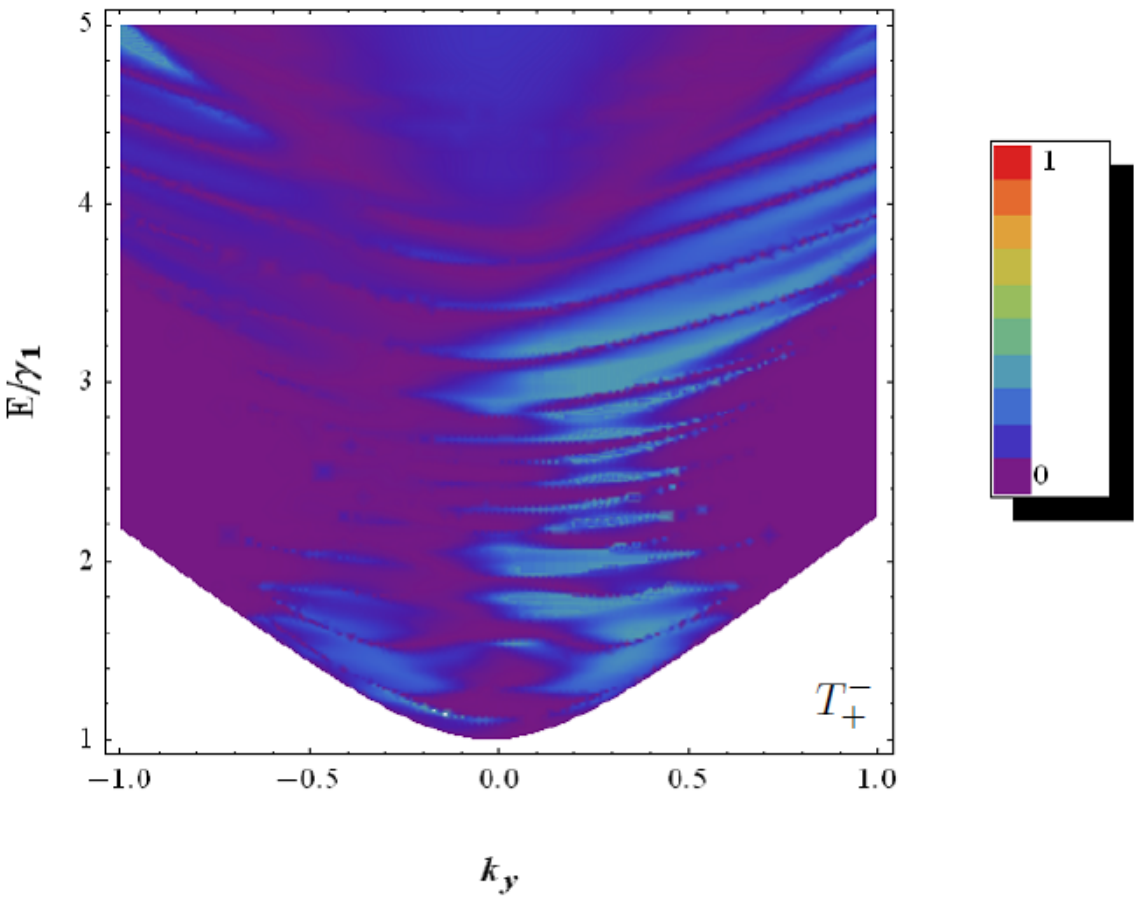}
  ~~~~~~ \includegraphics[width=6cm, height=5cm]{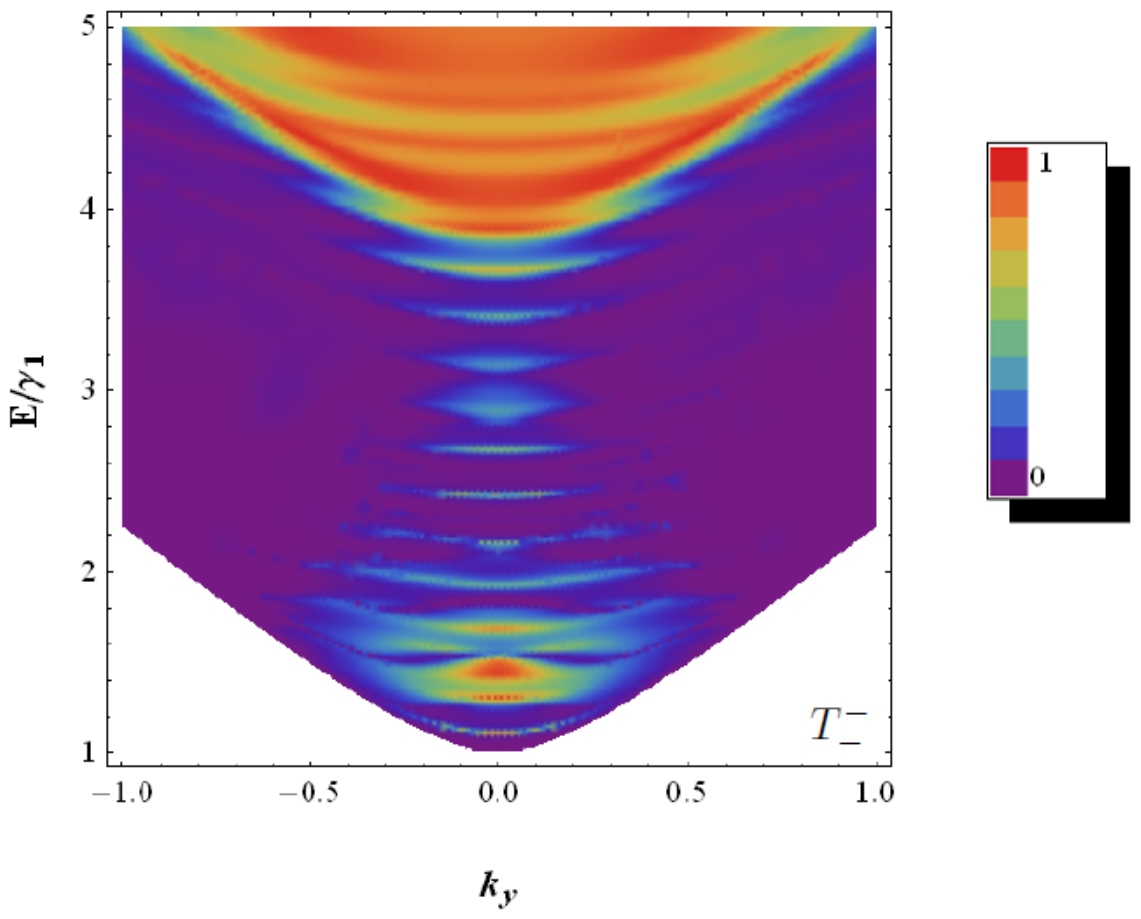}
 \caption{Density plot of the four transmission channels as a function of
the transverse wave vector $k_y$ and its energy $E$, with
$V_2=V_4=2.5\ \gamma_1$, $V_3=0\ \gamma_1$,
$\delta_2=\delta_4=0.4\ \gamma_1$, $\delta_3=0.\ \gamma_1$,
$l_B=18.5\ nm$, $d_4=-d_1=14\ nm$ and $d_3=-d_2=7.5\ nm$.}
\label{fig.FB2}
\end{figure}

In Figure \ref{fig.FB1} we show the different channels associated with
the four transmissions as a function of the transverse wave vector
$k_y$ and energy $E$. The height of the potential is set
to $V_2=V_4=2.5\ \gamma_1$ and $V_3=0\ \gamma_1$. We notice that the
Dirac fermions exhibit transmission resonance in $T_{+}^+$
channel in the region $E < V_2-\gamma_1$ where the electrons
propagate via $\alpha^+$. The cloak effect \cite{Gu} appears in the region
$ V_2-\gamma_1< E < V_2$ for nearly normal incidence in the
$T_{+}^+$, $T_{-}^{+}$ and $T_{+}^{-}$ channels, while for
non-normal incidence it does not appear \cite{Duppen,Hassan,Jellal}.
Introducing asymmetric double barrier structure in the presence
of the magnetic field will break this equivalence such
that $T_{-}^{+} \neq T_{+}^{-}$. In the four transmission channels
in the energy regime $E > V_2=V_4$ they are similar to those obtained in
Figure \ref{fig.FB3}. In the presence of the magnetic field and for the
energy regime $ E < V_2=V_4$, the resonances result from the
bound electrons. In addition, the resonant peaks in the regime of
energy $ E < V_2=V_4$ are more intense compared to those obtained in
Figure \ref{fig.FB3}. For $T_{-}^-$ electrons propagate via
$\alpha^-$ mode which is absent between the two barriers ($E <
V_2=V_4$) so that the transmission is suppressed in this region
and this is equivalent to the cloak effect \cite{Gu}.

Now let us investigate the effect of the interlayer potential
difference on the band structure. We show the four channels of the
transmission probabilities in Figure \ref{fig.FB2}. For the same
parameters as in Figure \ref{fig.FB1} but for an interlayer potential
difference $\delta_2=\delta_4=0.4\ \gamma_1$ and $\delta_3=0\
\gamma_1$.  The general behavior of these different channels
resembles the single barrier case in the presence of a magnetic
fields \cite{Jellal}, with some major differences such as the observation of
peaks in the region $V_2-\delta_2 < E < V_2+\delta_2$. These peaks are
due to the existence of bounded states \cite{Hassan}.

\section{Conclusion}
In summary, we evaluated the transmission probabilities in
AB-stacked bilayer graphene through a double barrier potential in
the presence of a uniform magnetic field. We formulated our model
Hamiltonian that describes the system and computed the associated energy
eigenvalues as well as the band structure. We obtained a full
four bands of the energy spectrum and solved for the spinor solution
in each region of our system. The boundary conditions were used to calculate the
transmission probabilities which were computed numerically.

We found that the transmission can be enhanced due to the presence of
two propagation modes whose energy scale is set by the interlayer
coupling. For energies less than the height of the barrier the Dirac
fermions exhibit transmission resonances and only one propagation mode
is available. The transmission at nearly normal incidence is zero and
does not show any resonances in the regime of energy $E < V_3$.
Resonances are present for non-normal incidence, this is equivalent
to the situation of a single barrier. In addition, the presence of
the peaks in the regime of energy $V_3 < E <V_2=V_4$ are due to the bound electron states.
However, these peaks are absent in the case of a single barrier. When
the energy is higher than the interlayer coupling two propagation
modes are available for transport giving rise to four possible
ways for transmission probabilities.

 We also found that for the
case with $V_3 \neq 0$ resonant peaks are more intense
compared to those obtained for $V_3 = 0$. Then, we studied how the
interlayer potential difference affects the transmission
probability. We found that the transmission displays sharp peaks
inside the transmission gap (around the Dirac point $E = V_2 =
V_4$), that are absent around $E = V_3$. These peaks can be
attributed to the bound states formed by the double barrier
potential.

\section*{Acknowledgments}
The authors would like to acknowledge the support of KFUPM under the group project
RG1306-1 and RG1306-2. The generous support provided by the Saudi Center for Theoretical
Physics (SCTP) is highly appreciated by all authors.


\begin{thebibliography}{99}


\bibitem{Geim07}           A. K. Geim and K.S. Novoselov, Nature Materials 6, 183 (2007).

\bibitem{Novo05}           K. S. Novoselov, A. K. Geim, S. V. Morozov, D. Jiang, M. I. Katsnelson, I. V. Grigorieva, S. V. Dubonos and A. A. Firsov, Nature 438, 197 (2005).
\bibitem{Zhang05}          Y. B. Zhang, Y. W. Tan, H. L. Stormer and P. Kim, Nature 438, 201 (2005).
\bibitem{Moro08}           S. V. Morozov, K. S. Novoselov, M. I. Katsnelson, F. Schedin, D. C. Elias, J. A. Jaszczak and A. K. Geim, Physical Review Letters 100, 016602 (2008).
\bibitem{Lin10}            Y. M. Lin, C. Dimitrakopoulos, K. A. Jenkins, D. B. Farmer, H. Y. Chiu, A. Grill and P. Avouris, Science 327, 662 (2010).
\bibitem{Castro09}         A. H. Castro Neto, F. Guinea, N. M. R. Peres, K. S. Novoselov and A. K. Geim, Reviews of Modern Physics 81, 109 (2009).

\bibitem{McCa06}           E. McCann and V. I. Fa{\l}ko, Physical Review Letters 96, 086805 (2006).
\bibitem{novo06}           K. S. Novoselov, E. McCann, S. V. Morozov, V. I. Fa{\l}ko, M. I. Katsnelson, U. Zeitler, D. Jiang, F. Schedin and A. Geim, Nature Physics 2, 177 (2006).

\bibitem{novo12}           K. S. Novoselov, V. I. Fa{\l}ko, L. Colombo, P. R. Gellert, M. G. Schwab and K. Kim, Nature 490, 192 (2012).

\bibitem{Zhou07}           S. Y. Zhou, D. A. Siegel, A. V. Fedorov, F. El Gabaly, A. K. Schmid, A. H. Castro Neto and A. Lanzara, Nature Materials 7, 259 (2007).
\bibitem{Costa07}          R. Costa Filho, G. Farias and F. Peeters, Physical Review B 76, 193409 (2007).
\bibitem{Zhang09}          Y. Zhang, T.-T. Tang, C. Girit, Z. Hao, M.C. Martin, A. Zettl, M.F. Crommie, Y.R. Shen and F. Wang, Nature 459, 820 (2009).
\bibitem{McCanB09}         E. McCann, Physical Review B 74, 1 (2006).

\bibitem{Guinea06}         F. Guinea , A. H. C. Neto and N. M. R. Peres, Physical Review B 73, 245426 (2006).
\bibitem{Latil06}          S. Latil and L. Henrard, Physical Review Letters 97, 036803 (2006).
\bibitem{Parto06}          B. Partoens and F. M. Peeters, Physical Review B 74, 075404 (2006).
\bibitem{Katsn06}          M. I. Katsnelson, K. S. Novoselov and A. K. Geim, Nature Physics 2, 620 (2006).
\bibitem{Duppen}           B. V. Duppen and F.M. Peeters, Physical Review B 87, 205427 (2013).
\bibitem{Hassan}           H. A. Alshehab, H. Bahlouli, A. El Mouhafid and A. Jellal, arXiv: 1401.5427 (2014).
\bibitem{Jellal}           A. Jellal, I. Redouani and H. Bahlouli, Physica E 72, 149 (2015) 

\bibitem{Wallace}          P. R. Wallace, Physical Review 71, 622 (1947); J. C. Slonczewski and P. R. Weiss, Physical Review 109, 272 (1958).
\bibitem{McClure}          J. W. McClure, Physical Review 108, 612 (1957).
\bibitem{Snyman}           I. Snyman and C. W. J. Beenakker, Physical Review B 75, 045322 (2007).
\bibitem{Gu}               N. Gu, M. Rudner and L. Levitov, Physical Review Letters 107, 156603 (2011).


\end{thebibliography}
\end{document}